# Tailoring dielectric resonator geometries for directional scattering and Huygens' metasurfaces

Salvatore Campione,[1,2,*] Lorena I. Basilio,[2] Larry K. Warne,[2] and Michael B. Sinclair[2]

[1]*Center for Integrated Nanotechnologies (CINT), Sandia National Laboratories, P.O. Box 5800, Albuquerque, NM 87185, USA*
[2]*Sandia National Laboratories, P.O. Box 5800, Albuquerque, NM 87185, USA*
[*]*sncampi@sandia.gov*

**Abstract:** In this paper we describe a methodology for tailoring the design of metamaterial dielectric resonators, which represent a promising path toward low-loss metamaterials at optical frequencies. We first describe a procedure to decompose the far field scattered by subwavelength resonators in terms of multipolar field components, providing explicit expressions for the multipolar far fields. We apply this formulation to confirm that an isolated high-permittivity dielectric cube resonator possesses frequency separated electric and magnetic dipole resonances, as well as a magnetic quadrupole resonance in close proximity to the electric dipole resonance. We then introduce multiple dielectric gaps to the resonator geometry in a manner suggested by perturbation theory, and demonstrate the ability to overlap the electric and magnetic dipole resonances, thereby enabling directional scattering by satisfying the first Kerker condition. We further demonstrate the ability to push the quadrupole resonance away from the degenerate dipole resonances to achieve local behavior. These properties are confirmed through the multipolar expansion and show that the use of geometries suggested by perturbation theory is a viable route to achieve purely dipole resonances for metamaterial applications such as wave-front manipulation with Huygens' metasurfaces. Our results are fully scalable across any frequency bands where high-permittivity dielectric materials are available, including microwave, THz, and infrared frequencies.

## 1. Introduction

Metallic resonators exhibit high intrinsic ohmic losses that preclude their use in resonant metamaterials operating at infrared and higher frequencies. Dielectric resonators represent a promising alternative building block for the development of low-loss resonant metamaterials because they replace lossy ohmic currents with low-loss displacement currents [1]. The spectral locations of electric and magnetic dipole resonances of a dielectric resonator can be tuned by varying the resonator geometry so that desired scattering properties are achieved. For example, by appropriately overlapping electric and magnetic dipole resonances, cancellation of scattering in backward or forward direction may be achieved as dictated by the Kerker conditions [2-4]. Assembling the resonators into two-dimensional periodic arrays may lead to similar behavior [5] (i.e., minima in reflection or transmission).

However, the use of dielectric resonators is not without its own challenges since achieving the desired resonant properties while maintaining a sufficiently small resonator size and spacing requires the use of very high permittivity materials. While permittivity values larger than 100 are readily available at THz and microwave frequencies, the largest permittivity currently available at infrared wavelengths is approximately 32 (e.g. lead telluride). Thus, the geometric details of the dielectric resonator design and their assembly into metamaterials are extremely constrained, and maintaining effective medium behavior is challenging. For this reason, the field of metamaterials has focused in recent years in the development of homogeneous artificial materials that are characterized by local effective material parameters [6-13]. A necessary condition for local behavior is that the metamaterial constituents possess only dominant (electric and/or magnetic) dipole resonances and negligible higher-order multipolar terms (e.g. quadrupoles, octupoles, etc.). This fact was very recently emphasized in [14] where the authors achieved *local magnetic metamaterials* through the use of the extreme coupling regime of cut-plate pairs or split ring resonators. In this paper we will make use of geometries based on perturbation theory, previously introduced in [15, 16], as an alternative route to obtain resonators that exhibit dominant dipole resonances in certain frequency bands. These perturbed resonators may then be used to achieve local properties in metamaterials. We further show that such perturbations of the resonator geometry provide additional degrees of freedom that allow us to overlap the electric and magnetic dipole resonances, possibly enabling negative-index- or zero-index-like functionalities. In [15, 16] the main objective was to introduce cavity-perturbation techniques to determine the types of inclusions (in terms of material, polarization, and placement) that are necessary to realize degenerate dipole resonances, as well as to derive simple formulas which can be used for the design of these types of resonators. Here, we start with the results in [15, 16] to investigate the use of such resonators for a directional scattering metamaterial application.

A common way to identify the multipoles that dominate the scattering response of isolated resonators is through the use of multipolar analysis or multipolar expansion [14, 17-23]. According to [17], the scattered field $\mathbf{E}_s$ produced by a sphere can in general be written as an infinite series in the vector spherical harmonics $\mathbf{N}_{emn}$ and $\mathbf{M}_{omn}$ (where the subscripts $e$ and $o$ stand for even and odd, respectively), the so-called electromagnetic *normal modes* of the spherical particle, weighted by appropriate coefficients $a_{mn}$ and $b_{mn}$ as

$$\mathbf{E}_s = \sum_{n=1}^{\infty} \sum_{m=-n}^{n} (a_{mn}\mathbf{N}_{emn} + b_{mn}\mathbf{M}_{omn}). \tag{1}$$

In Eq. (1), the index $n$ indicates the degree of the multipole (e.g. 1 = dipole, 2 = quadrupole, 3 = octupole, etc.) and $m$ indicates the possible orientations of the multipole. Equation (1) can be extended to model the scattered field produced by subwavelength resonators of any shape through suitable choice of the $a_{mn}$ and $b_{mn}$ coefficients. Although this is probably the most common multipolar expansion formulation due to its compactness and elegance, we prefer to express the multipolar components in terms of the multipole moments, e.g. $\mathbf{p}$, $\mathbf{m}$, and $\mathbf{Q}$ (their definitions will be provided in Sec. 2). This will give us a better insight on the far-field angular dependence otherwise hidden in the terms reported in Eq. (1).

In Sec. 2, and in the Appendix, we present explicit expressions for the multipole fields in terms of the multipole moments, and describe how the field scattered by an arbitrary (subwavelength) object can be decomposed into a sum of multipole fields. In Sec. 3 we apply this formulation to the case of a high-permittivity dielectric resonator that supports electric and magnetic dipole resonances in separate frequency bands, as well as a quadrupolar resonance, and show that the formulation clearly identifies the contribution of each multipole. We then modify the resonator geometry in a manner suggested by perturbation theory and show that this approach allows us to overlap the electric and magnetic dipole resonances, while simultaneously pushing away the quadrupolar resonance and thereby enabling local behavior at the dipole resonances. We also compute the electric and magnetic dipole polarizabilities of the perturbed resonators, and show that one may satisfy the first Kerker condition to obtain forward scattering behavior. In this regard, we demonstrate a metamaterial array of perturbed cubic resonators that exhibits high transmission and a $2\pi$ phase coverage — the characteristic properties required for high efficiency Huygens' metasurfaces [24, 25]. We envision that the additional degrees of freedom afforded by the perturbation approach will allow the design of resonators that are appealing for metamaterial applications [25, 26].

## 2. Theoretical framework of multipolar expansion

Consider the total far field $\mathbf{E}_{tot}$ scattered by a subwavelength resonator illuminated by a plane wave. As indicated by Eq. (1) and sketched in Fig. 1, the total far field can be decomposed in terms of multipolar components as [18]

$$\mathbf{E}_{tot} = \mathbf{E}_{ED} + \mathbf{E}_{MD} + \mathbf{E}_{EQ} + \mathbf{E}_{MQ} + \mathbf{E}_{EO} + \mathbf{E}_{MO} + \text{higher order terms}, \tag{2}$$

where the subscripts on the right hand side indicate electric and magnetic dipoles (ED and MD), electric and magnetic quadrupoles (EQ and MQ), and electric and magnetic octupoles (EO and MO). The list in Eq. (2) has been truncated purposely to the octupolar terms; higher-order terms are present, though they are negligible in most of the cases where the size of the scattering object is small compared to the wavelength [17].

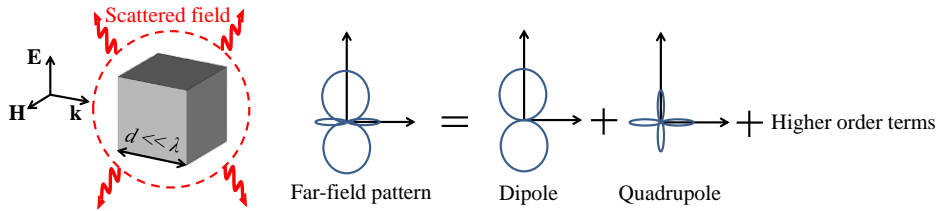

Fig. 1. A subwavelength resonator under plane wave illumination scatters a far field that can be decomposed in terms of multipolar field components, i.e. dipole, quadrupole, and higher order terms.

The contributions of the multipole components in Eq. (2) can be written in vectorial form as

$$\mathbf{E}_{ED} = Z_0 \frac{ck^2}{4\pi} \frac{e^{ikr}}{r} \hat{\mathbf{r}} \times \mathbf{p} \times \hat{\mathbf{r}}, \qquad \mathbf{E}_{MD} = -Z_0 \frac{k^2}{4\pi} \frac{e^{ikr}}{r} \hat{\mathbf{r}} \times \mathbf{m}, \qquad (3)$$

$$\mathbf{E}_{EQ} = -Z_0 \frac{ick^3}{24\pi} \frac{e^{ikr}}{r} \hat{\mathbf{r}} \times \mathbf{Q}_{EQ} \times \hat{\mathbf{r}}, \qquad \mathbf{E}_{MQ} = Z_0 \frac{ik^3}{24\pi} \frac{e^{ikr}}{r} \hat{\mathbf{r}} \times \mathbf{Q}_{MQ}, \qquad (4)$$

$$\mathbf{E}_{EO} = -Z_0 \frac{ck^4}{120\pi} \frac{e^{ikr}}{r} \hat{\mathbf{r}} \times \mathbf{O}_{EO} \times \hat{\mathbf{r}}, \qquad \mathbf{E}_{MO} = Z_0 \frac{k^4}{120\pi} \frac{e^{ikr}}{r} \hat{\mathbf{r}} \times \mathbf{O}_{MO}, \qquad (5)$$

where $\hat{\mathbf{r}} = \sin\theta\cos\phi\hat{\mathbf{x}} + \sin\theta\sin\phi\hat{\mathbf{y}} + \cos\theta\hat{\mathbf{z}}$ is the unit vector in the radial direction, $Z_0$ is the free-space wave impedance, $c$ is the speed of light, and $k = \omega/c$ is the free-space wavenumber, with $\omega$ the angular frequency. Moreover, $\mathbf{p}[\text{Cm}]$ is the electric dipole moment, $\mathbf{m}[\text{Am}^2]$ is the magnetic dipole moment, $\mathbf{Q}_{EQ}[\text{Cm}^2] = \underline{\underline{\mathbf{Q}}}_{EQ} \cdot \hat{\mathbf{r}}$ is the electric quadrupole moment, $\mathbf{Q}_{MQ}[\text{Am}^3] = \underline{\underline{\mathbf{Q}}}_{MQ} \cdot \hat{\mathbf{r}}$ is the magnetic quadrupole moment, $\mathbf{O}_{EO}[\text{Cm}^3] = (\underline{\underline{\mathbf{O}}}_{EO} \cdot \hat{\mathbf{r}}) \cdot \hat{\mathbf{r}}$ is the electric octupole moment, and $\mathbf{O}_{MO}[\text{Am}^4] = (\underline{\underline{\mathbf{O}}}_{MO} \cdot \hat{\mathbf{r}}) \cdot \hat{\mathbf{r}}$ is the magnetic octupole moment. Note that the terms $\underline{\underline{\mathbf{Q}}}_{EQ}$ and $\underline{\underline{\mathbf{Q}}}_{MQ}$ are symmetric tensors and traceless, i.e. $Q^{xx} + Q^{yy} + Q^{zz} = 0$, reducing the independent quadrupolar components to five [in agreement with Eq. (1) where $n = 2$, $m = -2, -1, 0, +1, +2$]. Similarly, $\underline{\underline{\mathbf{O}}}_{EO}$ and $\underline{\underline{\mathbf{O}}}_{MO}$ are symmetric tensors and traceless, i.e. $\sum_j O^{jji} = \sum_j O^{jij} = \sum_j O^{ijj} = 0$, reducing the independent octupolar components to seven [in agreement with Eq. (1) where $n = 3$, $m = -3, -2, -1, 0, +1, +2, +3$].

Explicit expressions for the far fields scattered by dipoles and quadrupoles are presented in the Appendix. In general, multipole (MP) far fields can be expressed as:

$$\mathbf{E}_{MP} = W_{MP} C_{MP} \mathbf{A}_{MP}(\theta, \phi) \qquad (6)$$

where $W_{MP}$ is the (complex) weight of the multipole moment and $C_{MP}$ is a radially dependent pre-factor given in the Appendix [e.g. $C_{MP} = Z_0 2ck^2 e^{ikr}/(3r)$ for an electric dipole]. $\mathbf{A}_{MP}(\theta, \phi)$ are the normalized angular functions that are contained in the curly braces in Eqs. (11) – (26). For example, $\mathbf{A}_{MP}(\theta, \phi) = (-\sin\phi\hat{\boldsymbol{\varphi}} + \cos\theta\cos\phi\hat{\boldsymbol{\theta}})/(8\pi/3)$ for an $x$-directed electric dipole. We exploit the orthonormality of these angular functions over the solid angle to extract the contribution of each multipole to the total scattered field [18]:

$$W_{MP} C_{MP} = \int_0^{2\pi}\int_0^{\pi} [\mathbf{E}_{tot} \cdot \mathbf{A}_{MP}(\theta, \phi)] \sin\theta d\theta d\phi. \qquad (7)$$

The total power associated with each multipole is then computed as

$$P_{MP} = \frac{|W_{MP}|^2 |C_{MP}|^2}{2Z_0}. \qquad (8)$$

In subsequent sections we will use the total radiated power to determine which multipoles make significant contributions to the overall field scattered by the dielectric resonators.

### 3. Multipolar expansion for a single dielectric cube under electric- and magnetic-field drive conditions

We consider electric- (E-) and magnetic- (H-) field drive conditions as sketched in Fig. 2 [27, 28]. These excitation schemes make use of two counter propagating plane waves to cancel either the magnetic field (for E-field drive) or the electric field (for H-field drive) at the center of the resonator. In this way, we can selectively excite either electric or magnetic resonances and locate their spectral locations independently, provided the scattering object is sufficiently subwavelength.

To demonstrate the utility of the multipole decomposition approach, we first analyze the simple case of a lead telluride (PbTe) dielectric cube with side $d = 1.53\,\mu\text{m}$ (about $1/7^{\text{th}}$ of the free-space wavelength at the magnetic resonance) and relative permittivity equal to $32.04 + i0.0566$ embedded in free space. Although for simplicity we consider resonators in free space, placement of the resonators on a layer of low-index materials such as barium fluoride (e.g., see [29]) may require minor modifications to the design but will not significantly alter the properties we discuss here. This resonator design leads to electric and magnetic resonances in the mid-infrared region of the spectrum. The scattered far field obtained from full-wave simulations is shown in Fig. 3(a) for two different sampling positions: 1) $\theta = 90°$ and $\phi = 90°$ for E-field drive; 2) $\theta = 90°$ and $\phi = 0°$ for H-field drive. In agreement with [16], we observe the magnetic dipole resonance at 28.31 THz (under H-field drive), the electric dipole resonance at 38.37 THz (under E-field drive), and the magnetic quadrupole resonance at 42.98 THz (under E-field drive). These resonances are explicitly marked in Fig. 3(a). Using the scattering cross sections in place of the radiated far-field amplitudes as for example done in [14] will lead to similar conclusions. To further validate our approach, we recalculate the results of Fig. 3(a) using the radiated powers of the three dominant multipoles ($m_{MD}^{y}$, $p_{ED}^{x}$, and $Q_{MQ}^{zy}$). The results are shown in Fig. 3(b), and very good agreement with Fig. 3(a) is observed.

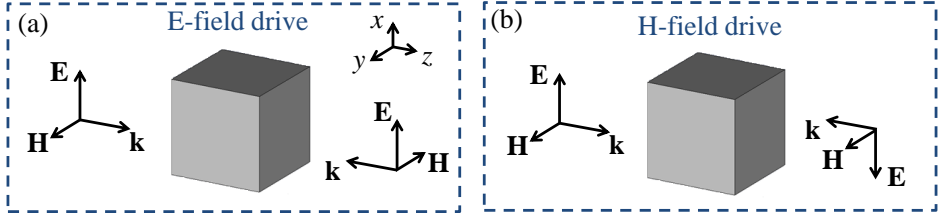

Fig. 2. Sketch of (a) E- and (b) H-field drive conditions. The phases of the counter propagating plane waves are chosen to cancel either the magnetic field (for E-field drive) or the electric field (for H-field drive) at the center of the resonator.

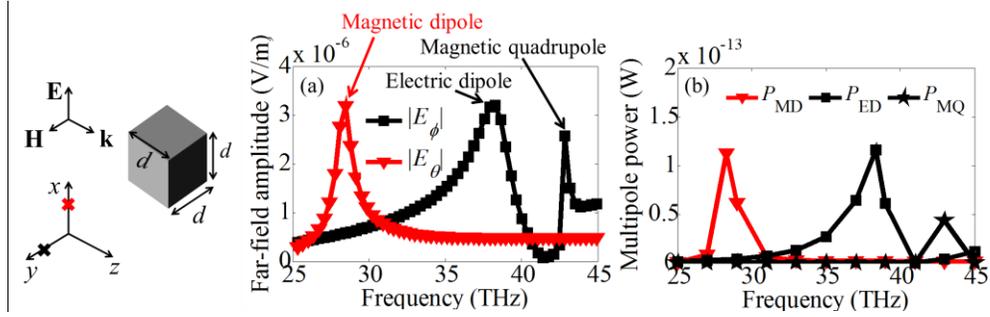

Fig. 3. (a) Radiated far-field amplitudes and (b) power associated with the multipoles $m_{MD}^{y}$, $p_{ED}^{x}$, and $Q_{MQ}^{zy}$ of a cubic dielectric resonator. Sampling positions are located on the $\theta = 90°$ plane at $\phi = 90°$ for E-field drive (black squares) and at $\phi = 0°$ for H-field drive (red triangles). The inset shows a schematic of the geometry including the sampling points depicted by black and red crosses.

Figure 4 compares the angular dependences of the far-field patterns at the three resonant frequencies obtained using full-wave simulations under E- and H-field drives to the fields obtained using the multipolar expansion methodology. Note that the multipole decomposition recovers the spectral and angular characteristics to a high degree of accuracy (the phase information is also recovered, although it is not shown for brevity). Table 1 summarizes the power associated with the dominant multipoles at the three resonant frequencies. As expected, the electric dipole and the magnetic dipole moments dominate at the electric and magnetic dipole resonance frequencies, respectively. At the magnetic quadrupole frequency, we find a dominating magnetic quadrupole moment along with a small (but not negligible) contribution of an electric dipole moment. The scattered powers of the multipoles not listed in Table 1 are smaller by at least two orders of magnitude. The scattered E-field patterns sampled at the three resonant frequencies are plotted in Fig. 5, where clear signatures of dipolar and quadrupolar fields are observed.

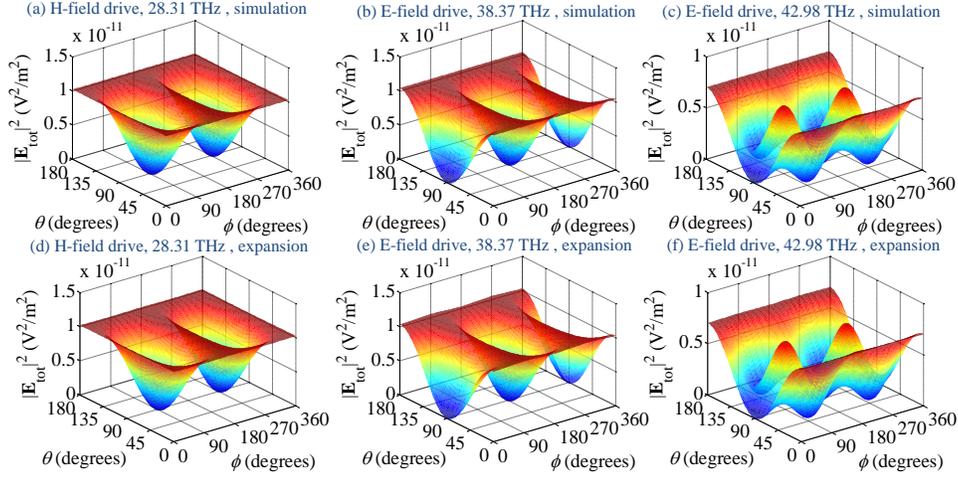

Fig. 4. Far-field patterns versus $\theta$ and $\phi$ for a cubic dielectric resonator at the magnetic dipole resonance, electric dipole resonance, and magnetic quadrupole resonance, computed via (a-c) full-wave simulations and reproduced via (d-f) multipolar expansion.

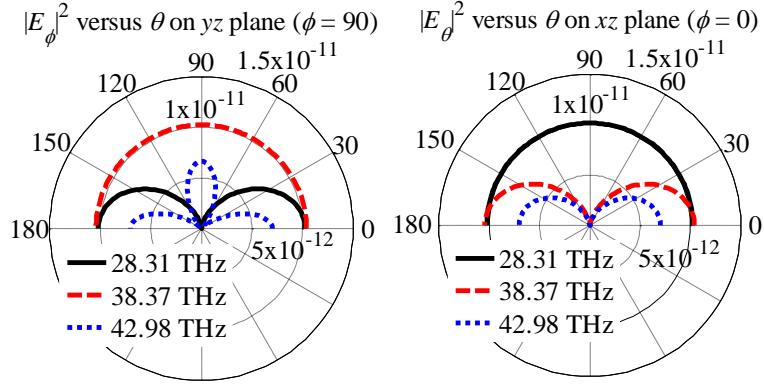

Fig. 5. Angular distribution in the *y-z* and *x-z* planes of the far-field scattered by a cubic dielectric resonator at the magnetic dipole resonance (black solid), electric dipole resonance (red dashed), and magnetic quadrupole resonance (blue dotted).

**Table 1. Power associated with each multipole for a subwavelength dielectric cube. The powers radiated by multipolar components not reported here are smaller by more than two orders of magnitude.**

| Frequency (THz) | 28.31 | 38.37 | 42.98 |
|---|---|---|---|
| Excitation scheme | H-field drive | E-field drive | E-field drive |
| Multipole | $m_{MD}^y$ | $p_{ED}^x$ | $p_{ED}^x$, $Q_{MQ}^{zy}$ |
| Power ($\times 10^{-13}$ W) | 1.13 | 1.16 | 0.04, 0.43 |

## 4. Overlapping the electric and magnetic dipole resonances with a single-split dielectric cube

As described in [15, 16], perturbation techniques can be used to obtain resonator geometries that selectively adjust the spectral locations of the resonances. These works suggested the use of split-cubes or split-spheres to overlap the electric and magnetic dipole resonances (this point will be further discussed in Sec. 6). The splits are arranged in such a manner as to selectively interact with the electric field pattern of the magnetic resonance and shift the resonance frequency upwards towards the electric resonance frequency.

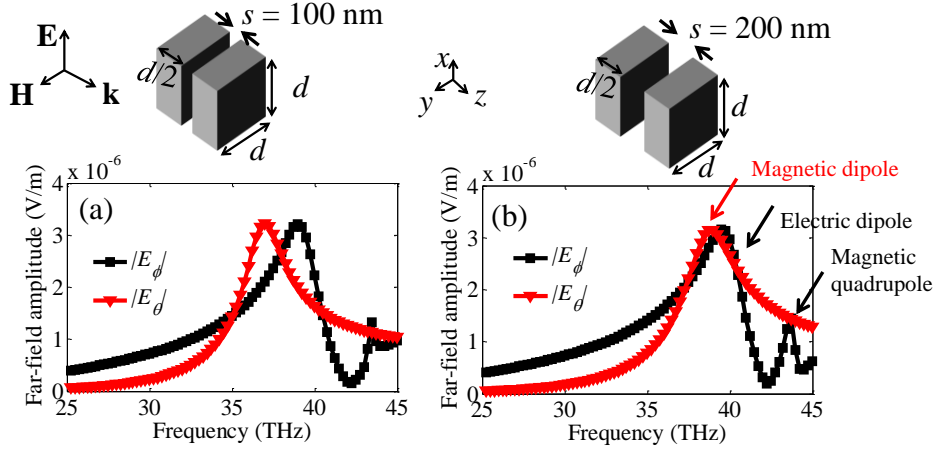

Fig. 6. Radiated far-field amplitudes of single-split cubes with gap of (a) 100 nm and (b) 200 nm. Sampling positions are located on the $\theta = 90°$ plane at $\phi = 90°$ for E-field drive (black squares) and at $\phi = 0°$ for H-field drive (red triangles). The insets show schematics of the two geometries.

We consider a dielectric cube containing a split in the midplane transverse to the plane wave propagation direction (creating a small gap between the two half cubes) as shown in the insets of Fig. 6. We analyze two values of the gap, namely 100 and 200 nm, to show that the perturbation causes the magnetic dipole resonance to move towards the electric one. We again perform full-wave simulations of the scattered field, and plot in Fig. 6 the radiated far-field amplitudes at two different sampling positions: 1) $\theta = 90°$ and $\phi = 90°$ for E-field drive; 2) $\theta = 90°$ and $\phi = 0°$ for H-field drive. Under H-field drive we observe the magnetic dipole resonance at 36.97 and 38.97 THz for 100 and 200 nm single-split cubes, respectively. With E-field drive we observe the electric dipole resonance at 38.97 and 39.47 THz and the magnetic quadrupole resonance at 43.47 and 43.72 THz for the 100 and 200 nm single-split cubes, respectively. These resonances are explicitly marked in Fig. 6(b). For the 200 nm single-split cube of Fig. 6(b), the introduction of the 200 nm gap raises the frequency of the magnetic dipole resonance to overlap that of the electric dipole resonance (the resonator size is about 1/5$^{th}$ of the free-space wavelength at the magnetic resonance). In contrast, the frequencies of the electric dipole and magnetic quadrupole resonances were largely unaffected by the introduction of the split. Once again, the multipole decomposition accurately recovers all the spectral, angular, and phase characteristics of the scattered fields (not shown). Table 2 summarizes the power radiated by each of the dominant multipoles. The powers radiated by multipoles not listed in Table 2 were smaller by at least two orders of magnitude. Note that both quadrupolar and electric dipolar behaviors were observed at the magnetic quadrupole resonance.

The introduction of splits decreases the symmetry of the cubes and it becomes convenient to describe the dipole moments in terms of electric and magnetic dipole polarizability tensors defined through [30]:

$$\mathbf{p} = \underline{\boldsymbol{\alpha}}_{ee} \cdot \mathbf{E}_{loc}, \quad \mathbf{m} = \underline{\boldsymbol{\alpha}}_{mm} \cdot \mathbf{H}_{loc} \qquad (9)$$

where $\underline{\boldsymbol{\alpha}}_{ee}$ and $\underline{\boldsymbol{\alpha}}_{mm}$ are the electric and magnetic dipole polarizability tensors and $\mathbf{E}_{loc}$ and $\mathbf{H}_{loc}$ are the local electric and magnetic fields acting on the resonator. For isotropic resonators, the polarizability tensors will be diagonal with equal components. The polarizability tensor of the split cubes will be diagonal in our scattering geometry, however some components will be different from each other. For this reason, we show in the following only the transverse components, here marked simply as $\alpha_{ee}$ and $\alpha_{mm}$.

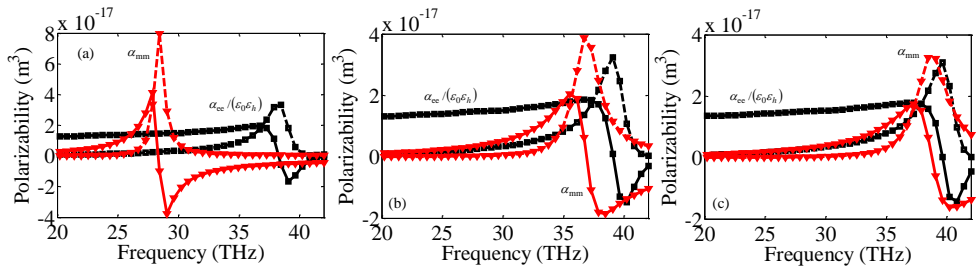

Fig. 7. The relative location of electric (black squares) and magnetic (red triangles) polarizabilities of subwavelength resonators is controllable through geometry. Solid: Real part; dashed: imaginary part. (a) Full-cube. (b) Single-split cube with gap $s$ = 100 nm. (c) Single-split cube with gap $s$ = 200 nm and $d$ = 1.53 μm. The monochromatic time harmonic convention, $\exp(-i\omega t)$, is assumed.

By following the multipolar decomposition procedure described above, we are able to estimate the electric and magnetic dipole polarizabilities of full cubes and single-split cubes. Figure 7 shows the results of the decomposition procedure for a full cube as well as single-split cubes with two different gap widths. Note that this figure presents polarizabilities in units of m$^3$, i.e. the electric polarizability in Eq. (9) is normalized to the host absolute permittivity $\varepsilon_0\varepsilon_h$. For the full cube we observe a magnetic dipole resonance around 28.31 THz (under H-field drive) followed by an electric dipole resonance at about 38.37 THz (under E-field drive), in agreement with the results in Sec. 2. The simulation results of Figs. 7(b)-7(c) are also in agreement with the results shown in Fig. 6, and show that the introduction of the split causes the magnetic resonance to move toward higher frequencies (red triangles), while leaving the electric resonance frequency unaffected (black squares). For the 200 nm single-split cube of Fig. 7(c), the magnetic dipole resonance (38.97 THz) is nearly overlapped with the electric dipole resonance (39.47 THz).

Table 2. Powers radiated by the dominant multipoles for a subwavelength single-split dielectric cube ($s$ = 200 nm and $d$ = 1.53 μm). The powers radiated by multipolar components not reported here are smaller by more than two orders of magnitude.

| Frequency (THz) | 38.97 | 39.47 | 43.72 |
|---|---|---|---|
| Excitation scheme | H-field drive | E-field drive | E-field drive |
| Multipole | $m_{MD}^y$ | $p_{ED}^x$ | $p_{ED}^x$, $Q_{MQ}^{zy}$ |
| Power ($\times 10^{-13}$ W) | 1.36 | 1.29 | 0.066, 0.03 |

As mentioned in Sec. 1, directional forward or backward scattering for isolated resonators can be obtained by appropriately overlapping electric and magnetic resonances [2-4]. In particular, the first Kerker condition [2, 3] states that the isolated resonator will predominantly scatter light in the forward direction when the Mie electric and magnetic dipole coefficients are equal ($a_1 = b_1$) and significantly larger than any higher order Mie terms $a_n, b_n : n > 1$. These conditions can be equivalently expressed through the electric and magnetic dipole polarizabilities as $\alpha_{ee}/(\varepsilon_0\varepsilon_h) = \alpha_{mm}$, since $\alpha_{ee}/(\varepsilon_0\varepsilon_h) = 6\pi i a_1/k^3$ and $\alpha_{mm} = 6\pi i b_1/k^3$. Interestingly, we observe in Fig. 8(a) that this condition is (almost perfectly) satisfied for both real and imaginary components near 37.29 and 40.82 THz (indicated by the vertical dashed-dotted orange lines). Figure 8(b) shows the scattered radiation pattern of an isolated single-split dielectric cube resonator (gap of 200 nm) excited through plane wave incidence for three excitation frequencies: 25, 37.29, and 40.82 THz. As expected, a single-lobed radiation pattern – a signature of forward scattering – is obtained at the two frequencies that satisfy the Kerker condition (37.29 and 40.82 THz). In contrast, a weaker, two-lobed scattering pattern is observed at the nonresonant frequency of 25 THz for which the Kerker condition is not satisfied. We envision that it may be possible to satisfy the Kerker conditions over a frequency band, rather than at isolated frequencies, by further tailoring the resonator design to better overlap the spectral position, width, and amplitude of the two polarizabilities.

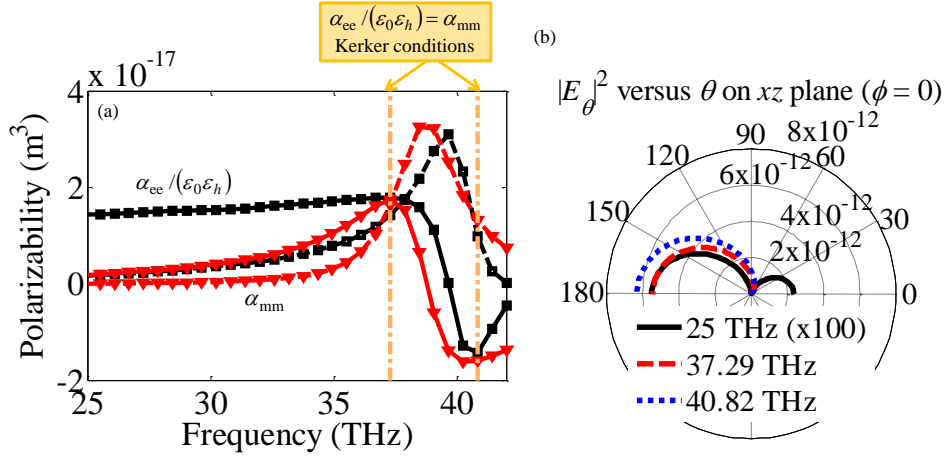

Fig. 8. (a) Polarizability result shown in Fig. 6(b). The frequencies that satisfy the first Kerker condition are indicated by the dashed-dotted vertical lines. (b) Scattered radiation pattern of an isolated single-split dielectric cube resonator excited through plane wave incidence for three excitation frequencies: forward scattering is evident (i.e. only one lobe at $\theta = 180$ degrees) when the first Kerker condition is satisfied. Only data between 0 and 180 degrees is reported; the scattering is specular between 180 and 360 degrees.

Until now we have focused on the multipolar characteristics of isolated dielectric resonators. To assess the applicability of these resonator geometries to metamaterials, consider a two-dimensional array of resonators arranged on a square lattice with a period of 2.6 μm. The reflectance and transmittance under normal plane wave incidence for the resonator geometries of Fig. 7(a) (full cube) and Fig. 7(c) (single-split cube with 200 nm gap) are shown in Figs. 9(a)-9(b). We observe fundamental differences between the spectra obtained for the two geometries. The array of full dielectric cubes (which possess electric and magnetic dipole resonances in separate frequencies) exhibits two strong reflection maxima and two corresponding transmission minima. In contrast, the array of split-cubes is highly transmissive over a wide frequency band because of the near overlap of dipolar resonances as shown in Fig. 7(c) and Fig. 8(a). These results are in accord with those shown in [5] for an array of silicon nanocylinders. The phases of the reflection and transmission coefficients at a distance of 8 μm from the array plane are shown in Figs. 9(c)-9(d) for the two resonator geometries. In agreement with [24], the transmission coefficient for the full cube metamaterial undergoes a phase shift of at most 180 degrees at each resonance, while the transmission coefficient for the split-cube metamaterial undergoes a complete 360 degree phase shift. This combination of features — high transmittance and 360 degree phase shift — renders the split-cube metamaterial design appealing for use in Huygens' metasurfaces [24] which are a promising platform for the development of flat optical devices [31, 32]. The reflection behavior is opposite that of the transmission: high reflectivity across a broad spectral range and a full 360 degree phase shift are only obtained for the full cube metamaterial. Thus, the full cube metamaterial allows us to manipulate the reflection response and realize a metareflector that can be very easily fabricated. These profound differences between the behaviors of the two metamaterials demonstrate the dramatic impact that perturbations at the single resonator level can have on the metamaterial performance at the macroscopic level.

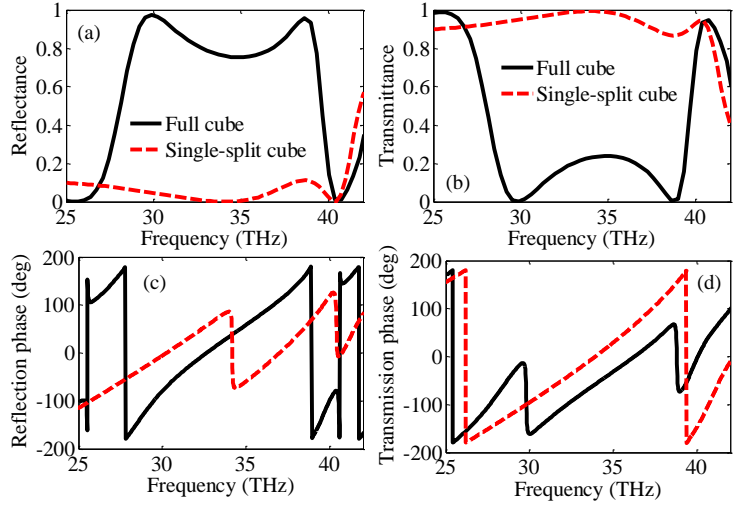

Fig. 9. (a) Reflectance and (b) transmittance of a two-dimensional array of dielectric resonators [full cubes as in Fig. 6(a) and single-split cubes as in Fig. 6(c)] arrayed on a square lattice with a period of 2.6 µm. Phase of the (c) reflection coefficient and of the (d) transmission coefficient for the cases in (a)-(b).

## 5. Pushing the quadrupolar resonance away from the overlapping electric and magnetic dipole resonances with a four-split dielectric cube

We now analyze the dielectric cube resonator shown in the inset of Fig. 10 which includes four-splits of width $s =$ 50 nm and an overall width of $d + s + s\sqrt{2}$ (the resonator is about $1/5^{\text{th}}$ of the free-space wavelength at the magnetic resonance). Following the same procedure used to retrieve Fig. 3, we observe that under H-field drive this resonator exhibits a magnetic dipole resonance at 37.59 THz. Under E-field drive an electric dipole resonance at 39.34 THz and a magnetic quadrupole resonance at 46.69 THz are observed as shown in Fig. 10. Once again, the multipole decomposition accurately recovers all the spectral, angular, and phase characteristics of the scattered far fields (not shown). Table 3 summarizes the power radiated by each of the dominant multipoles. Thus, in addition to nearly overlapping the electric and magnetic dipole resonances, the use of multiple splits pushes the quadrupolar resonance to higher frequency (see Sec. 6), enabling the development of local metamaterial properties.

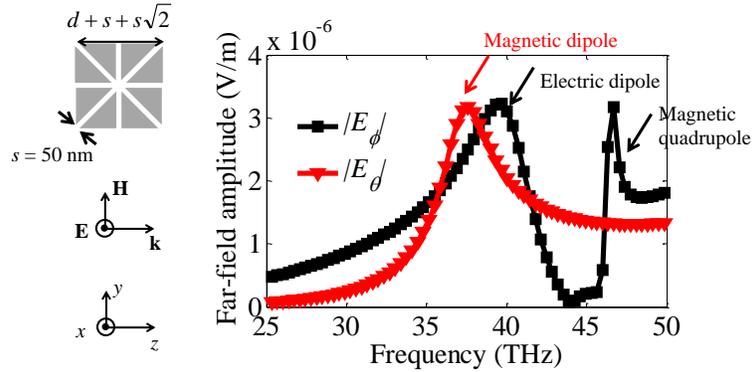

Fig. 10. Radiated far-field amplitudes of a four-split cube as in the inset. Sampling positions are located on the $\theta = 90°$ plane at $\phi = 90°$ for E-field drive (black squares) and at $\phi = 0°$ for H-field drive (red triangles).

Table 3. Powers radiated by the dominant multipoles for a subwavelength four-split dielectric cube ($s = 50$ nm and $d = 1.53$ µm). The powers radiated by multipolar components not reported here are smaller by more than two orders of magnitude.

| Frequency (THz) | 37.59 | 39.34 | 46.69 |
|---|---|---|---|
| Excitation scheme | H-field drive | E-field drive | E-field drive |
| Multipole | $m_{MD}^{y}$ | $p_{ED}^{x}$ | $p_{ED}^{x}$, $Q_{MQ}^{zy}$ |
| Power ($\times 10^{-13}$ W) | 1.38 | 1.18 | 0.11, 0.35 |

## 6. Discussion of perturbation theory – potential and outlook

In this paper we have presented several examples that highlight the potential for using perturbation theory [15, 16] to obtain resonator geometries that selectively adjust the spectral locations of the resonances to achieve desired metamaterial properties including local behavior. We have used single or multiple thin splits to overlap the magnetic and electric dipole resonances by upshifting the frequency of the magnetic dipole to that of the electric dipole. The use of multiple splits also has the further advantage of moving the quadrupole resonance away from the dipole resonances (which has the potential of lowering losses and enabling local behavior for metamaterial applications). To better visualize this property, we define the percentage quadrupolar resonance shift as

$$\text{Quadrupolar resonance shift}(\%) = \frac{f_Q - f_D}{f_D} \times 100 \qquad (10)$$

where $f_Q$ is the frequency of the quadrupole resonance, and $f_D$ is the frequency of the closest dipole resonance. The frequency location of the quadrupolar resonance is shown in Fig. 11(a), and the quadrupolar resonance shift given by Eq. (10) is reported as a bar diagram in Fig. 11(b) for the three resonator shapes analyzed in this paper [we note again that the three designs have almost equal electric dipole resonance frequency location, as can be observed in Fig. 11(a)]. It is evident that the four-split cube design has nearly doubled the quadrupolar resonance shift with respect to the single-split cube design, while keeping electric and magnetic dipolar resonances in the same spectral region.

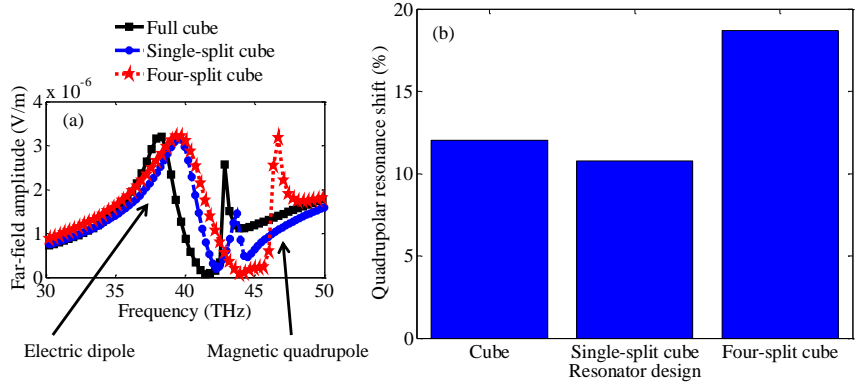

Fig. 11. (a) Radiated far-field amplitudes ($|E_\phi|$) of the three resonator designs analyzed in this paper. Sampling positions are located on the $\theta = 90°$ plane at $\phi = 90°$ for E-field drive. (b) Quadrupolar resonance shift in Eq. (10) versus the three resonator designs analyzed in this paper.

Alternatively, one could realize a frequency downshift of the electric dipole resonance toward the magnetic dipole resonance by embedding a metallic dipole, oriented along the direction of the incident electric field, within the dielectric resonator [15, 16]. This approach could however increase the losses and be counterproductive if employed at infrared or higher frequencies.

In these two alternative approaches, only one perturbation type was employed to selectively frequency shift one resonance while leaving the other unperturbed. It is natural to conclude that these approaches can be combined to enable operation away from the resonant peaks to overcome frequency up/downshift saturation, to allow smaller individual inclusions to be used, and to make the resonator design somewhat invariant with respect to incident plane-wave angle. For example, if minimizing the size of the perturbations is of interest, perturbations can be combined so as to simultaneously frequency shift the magnetic and electric dipole resonances toward each other and ultimately realize overlap at some intermediate frequency with respect to the fundamental ones [15, 16]. Advantages of the dual-perturbation design include the possibility of achieving resonator electrical sizes smaller than the single-split design, as well as the circumvention of frequency up/downshift saturation effects.

## 7. Conclusion

The design methodology presented in this paper affords great flexibility for tailoring the properties of dielectric resonators while also maintaining the subwavelength geometries required for local metamaterial properties. In particular, we employ perturbation theory to dielectric resonators to control the spectral overlap of electric and magnetic dipole resonances, as well as the location of higher-order modes. We make use of the multipolar expansion to confirm such properties and show that perturbation theory is a viable route to achieve purely dipole resonances in relatively wide frequency bands useful for the development of metamaterial applications. We show as a representative example that we can design a Huygens' metasurface that might enable engineerable wave-front control.

**Appendix: Explicit expressions of multipolar far fields up to the quadrupolar terms**

We report here the explicit far-field expressions up to the quadrupolar terms. The electric dipole has three independent components, namely $p_{ED}^x$, $p_{ED}^y$, and $p_{ED}^z$. Their far-field radiation is given by

$$\mathbf{E}_{ED} = Z_0 \frac{2ck^2}{3} \frac{e^{ikr}}{r} p_{ED}^x \left\{ \frac{-\sin\phi \hat{\boldsymbol{\phi}} + \cos\theta \cos\phi \hat{\boldsymbol{\theta}}}{8\pi/3} \right\} \tag{11}$$

$$\mathbf{E}_{ED} = Z_0 \frac{2ck^2}{3} \frac{e^{ikr}}{r} p_{ED}^y \left\{ \frac{\cos\phi \hat{\boldsymbol{\phi}} + \cos\theta \sin\phi \hat{\boldsymbol{\theta}}}{8\pi/3} \right\} \tag{12}$$

$$\mathbf{E}_{ED} = Z_0 \frac{2ck^2}{3} \frac{e^{ikr}}{r} p_{ED}^z \left\{ \frac{-\sin\theta \hat{\boldsymbol{\theta}}}{8\pi/3} \right\} \tag{13}$$

Similarly, the magnetic dipole has three independent components, namely $m_{MD}^x$, $m_{MD}^y$, and $m_{MD}^z$, whose far-field radiation is given by

$$\mathbf{E}_{MD} = -Z_0 \frac{2k^2}{3} \frac{e^{ikr}}{r} m_{MD}^x \left[ \frac{\cos\theta \cos\phi \hat{\boldsymbol{\phi}} + \sin\phi \hat{\boldsymbol{\theta}}}{8\pi/3} \right] \tag{14}$$

$$\mathbf{E}_{MD} = -Z_0 \frac{2k^2}{3} \frac{e^{ikr}}{r} m_{MD}^y \left[ \frac{\cos\theta \sin\phi \hat{\boldsymbol{\phi}} - \cos\phi \hat{\boldsymbol{\theta}}}{8\pi/3} \right] \tag{15}$$

$$\mathbf{E}_{MD} = -Z_0 \frac{2k^2}{3} \frac{e^{ikr}}{r} m_{MD}^z \left[ \frac{-\sin\theta \hat{\boldsymbol{\phi}}}{8\pi/3} \right] \tag{16}$$

The electric quadrupole has five independent components, namely $Q_{EQ}^{xx}$, $Q_{EQ}^{yx}$, $Q_{EQ}^{zx}$, $Q_{EQ}^{yy}$, and $Q_{EQ}^{zy}$. Their far-field radiation is given by

$$\mathbf{E}_{EQ} = -Z_0 \frac{ick^3}{15} \frac{e^{ikr}}{r} Q_{EQ}^{xx} \left[ \frac{-\sin\theta \sin\phi \cos\phi \hat{\boldsymbol{\phi}} + \sin\theta \cos\theta (1+\cos^2\phi) \hat{\boldsymbol{\theta}}}{8\pi/5} \right] \tag{17}$$

$$\mathbf{E}_{EQ} = -Z_0 \frac{ick^3}{15} \frac{e^{ikr}}{r} Q_{EQ}^{yx} \left[ \frac{\sin\theta (2\cos^2\phi - 1) \hat{\boldsymbol{\phi}} + 2\sin\theta \cos\theta \sin\phi \cos\phi \hat{\boldsymbol{\theta}}}{8\pi/5} \right] \tag{18}$$

$$\mathbf{E}_{EQ} = -Z_0 \frac{ick^3}{15} \frac{e^{ikr}}{r} Q_{EQ}^{zx} \left[ \frac{-\sin\phi \cos\theta \hat{\boldsymbol{\phi}} + (2\cos^2\theta - 1)\cos\phi \hat{\boldsymbol{\theta}}}{8\pi/5} \right] \tag{19}$$

$$\mathbf{E}_{EQ} = -Z_0 \frac{ick^3}{15} \frac{e^{ikr}}{r} Q_{EQ}^{yy} \left[ \frac{\sin\theta \sin\phi \cos\phi \hat{\boldsymbol{\phi}} + \sin\theta \cos\theta (1+\sin^2\phi) \hat{\boldsymbol{\theta}}}{8\pi/5} \right] \tag{20}$$

$$\mathbf{E}_{EQ} = -Z_0 \frac{ick^3}{15} \frac{e^{ikr}}{r} Q_{EQ}^{zy} \left[ \frac{\cos\theta\cos\phi\hat{\boldsymbol{\varphi}} + (2\cos^2\theta - 1)\sin\phi\hat{\boldsymbol{\theta}}}{8\pi/5} \right] \quad (21)$$

The magnetic quadrupole has five independent components, namely $Q_{MQ}^{xx}$, $Q_{MQ}^{yx}$, $Q_{MQ}^{zx}$, $Q_{MQ}^{yy}$, and $Q_{MQ}^{zy}$. Their far-field radiation is given by

$$\mathbf{E}_{MQ} = Z_0 \frac{ik^3}{15} \frac{e^{ikr}}{r} Q_{MQ}^{xx} \left[ \frac{\sin\theta\cos\theta(1+\cos^2\phi)\hat{\boldsymbol{\varphi}} + \sin\theta\sin\phi\cos\phi\hat{\boldsymbol{\theta}}}{8\pi/5} \right] \quad (22)$$

$$\mathbf{E}_{MQ} = Z_0 \frac{ik^3}{15} \frac{e^{ikr}}{r} Q_{MQ}^{yx} \left[ \frac{2\sin\theta\cos\theta\sin\phi\cos\phi\hat{\boldsymbol{\varphi}} + \sin\theta(1-2\cos^2\phi)\hat{\boldsymbol{\theta}}}{8\pi/5} \right] \quad (23)$$

$$\mathbf{E}_{MQ} = Z_0 \frac{ik^3}{15} \frac{e^{ikr}}{r} Q_{MQ}^{zx} \left[ \frac{\cos\phi(2\cos^2\theta - 1)\hat{\boldsymbol{\varphi}} + \cos\theta\sin\phi\hat{\boldsymbol{\theta}}}{8\pi/5} \right] \quad (24)$$

$$\mathbf{E}_{MQ} = Z_0 \frac{ik^3}{15} \frac{e^{ikr}}{r} Q_{MQ}^{yy} \left[ \frac{\sin\theta\cos\theta(1+\sin^2\phi)\hat{\boldsymbol{\varphi}} - \sin\theta\sin\phi\cos\phi\hat{\boldsymbol{\theta}}}{8\pi/5} \right] \quad (25)$$

$$\mathbf{E}_{MQ} = Z_0 \frac{ik^3}{15} \frac{e^{ikr}}{r} Q_{MQ}^{zy} \left[ \frac{\sin\phi(2\cos^2\theta - 1)\hat{\boldsymbol{\varphi}} - \cos\theta\cos\phi\hat{\boldsymbol{\theta}}}{8\pi/5} \right] \quad (26)$$

**Acknowledgments**

The authors acknowledge fruitful discussions with Dr. O. Wolf, Sandia National Laboratories. This work was performed, in part, at the Center for Integrated Nanotechnologies, a U.S. Department of Energy, Office of Basic Energy Sciences user facility. Portions of this work were supported by the Laboratory Directed Research and Development program at Sandia National Laboratories. Sandia National Laboratories is a multi-program laboratory managed and operated by Sandia Corporation, a wholly owned subsidiary of Lockheed Martin Corporation, for the U.S. Department of Energy's National Nuclear Security Administration under contract DE-AC04-94AL85000.